\documentclass[runningheads]{llncs}
\usepackage{listings}
\usepackage{graphicx}
\usepackage[belowskip=-5pt,aboveskip=0pt]{caption}
\usepackage{threeparttable}

\usepackage{subcaption}
\makeatletter
\def\thanks#1{\protected@xdef\@thanks{\@thanks
        \protect\footnotetext{#1}}}

\begin{document}
\title{TopicsRanksDC: Distance-based Topic Ranking applied on Two-Class Data\thanks{This work has been partially supported by the "Wachstumskern Qurator – Corporate Smart Insights" project (03WKDA1F) funded by the German Federal Ministry of Education and Research (BMBF)}}
\author{Malik Yousef\inst{1} \and
Jamal Al Qundus\inst{2} \and
Silvio Peikert\inst{2}\and
Adrian Paschke\inst{2}}
\authorrunning{M. Yousef et al.}
\institute{Zefat Academic College, Zefat,Israel
\email{malik.yousef@gmail.com}\\
The Galilee Digital Health Research Center (GDH),Zefat
%\url{http://www.springer.com/gp/computer-science/lncs} 
\and
Data Analytics Center (DANA), Fraunhofer FOKUS, Berlin\\
\email{\{jamal.al.qundus,silvio.peikert,adrian.paschke\}@fokus.fraunhofer.de}}
\maketitle              % typeset the header of the contribution
\begin{abstract}
In this paper, we introduce a novel approach named TopicsRanksDC for topics ranking based on the distance between two clusters that are generated by each topic. We assume that our data consists of text documents that are associated with two-classes. Our approach ranks each topic contained in these text documents by its significance for separating the two-classes. Firstly, the algorithm detects topics using Latent Dirichlet Allocation (LDA). The words defining each topic are represented as two clusters, where each one is associated with one of the classes. We compute four distance metrics, Single Linkage, Complete Linkage, Average Linkage and distance between the centroid. We compare the results of LDA topics and random topics. The results show that the rank for LDA topics is much higher than random topics. The results of TopicsRanksDC tool are promising for future work to enable search engines to suggest related topics.

\end{abstract}
\section{Introduction}
Information is regularly digitally generated and archived. Searching for documents in digital archives becomes more and more difficult over time, and parallel to this, the number of use cases and their interrelationships that need to be covered by information retrieval systems are growing. In this sense, there is a constant need for techniques that help us to organize, explore and understand data.
One important technique is topic modeling that performs analysis on texts to identify topics. These topic models are used to classify documents and to support further algorithms to perform context adaptive feature, fact and relation extraction \cite{1_alqundus_ai_supported}. While Latent Dirichlet Allocation (LDA)\cite{2_blei_lda}, Pachinko Allocation \cite{3_wei_dag_structured}, or Probabilistic Latent Semantic Analysis (PLSA) \cite{4_hofmann_plsi} traditionally perform topic modeling by statistical analysis of co-occurring words, the approaches \cite{5_allahyari_automatic} \cite{6_hulpus_unsupervised} in  integrate semantics into LDA. In general, topic modeling is about capturing the relationships among the words and documents (in terms of topics), and calculating the likelihood of belonging a word to a topic.
However, in many situations when working with large amounts of data, it is very difficult to help the user to quickly grasp a text collection or to get an overview of identified topics. Therefore, it is helpful to rank all topics and focus on the relevant ones that are both pressing and significant.

This is precisely the aim of topic ranking approaches, namely not to offer all identified topics in the same way, but to investigate the correlation of topics that correspond to a given sector and to present those with higher priority.
Ranking of topics could lead to a loss of information, which occurs when relevant topics or their clusters are ranked too low and are therefore no longer represented in the ranking. This problem can be alleviated by merging corresponding or similar clusters. This leads to the research question: 
To what extent can measuring the cluster distance of topics support the topic ranking?
The intuitive and basic principle for calculating distances fits the Euclidean equation:
$d_{euclidean}(x , y) = \sqrt{\sum_{i=1}^{n}(y_i-x_i)^2}$.
The challenge is to select the appropriate representative points of the clusters. This selection depends on the desired technique for measuring distance. In the case of a single-linkage, these points are the closest between the two clusters. In contrast, Complete-link distance uses the points furthest from the other cluster. While the average-link applies the average distance between all points of both clusters, otherwise, the distance of the centroids the clusters can be considered. This paper aims to determine the degree of likeness of two clusters, putting the focus on the minimum distance between the clusters and thus make use of the single- linkage method.
The paper is structured as follows: Section II provides a brief overview of relevant works. Section III describes. Section IV presents the preparation. Section VII reports and discusses the findings. Section VIII contains the summary and concludes with proposals for further investigation.

\section{Related Work}\label{sec:relatedwork}

Topic Modeling and Ranking are very popular and in great demand. Several previous studies have investigated these topics and their development, which is also being followed intensively by industry and scientists. The methods used are basically comparable, but differ in terms of the objectives, such as reducing the number of insignificant, similar or even widely divergent topics. It should be noted that summary does not always have to be the result of a topic-relevant search, as topics can never match 100\%. The study of \cite{7_alsumait_topic_significance} presents a semantic-based approach for the automatic classification of LDA topics to finally identify junk and insignificant topics. The authors measure how "different" the distribution of topics is as a "junk" distribution and thus the degree of insignificance a derived topic carries in its distribution. According to the Zipf law\footnote{Joachims, T.: A Statistical Learning Model of Text Classification with Support Vector Machines. In: Proceedings of the Conference on Research and Development in Information Retrieval, SIGIR (2001)} : A real topic can be modelled as a distribution of small number of words, so-called "salient words". Conversely, the "junk distribution" means that a large number of terms probably represent insignificant topics. 
Kullback-Leibler (KL)\footnote{Bishop, C.M.: Pattern Recognition and Machine Learning. Springer, Heidelberg (2006)} divergence was used to calculate the distance of the topic distribution over the number of salient words. Unlike our work, this work is based on an unsupervised quantification of topic meaning by identifying junk and insignificant topics. While our work aims to find relevant topics. The study of \cite{8_song_topic_and_keyword} proposes a method for re-ranking topics, motivated by the observation that similar documents are likely to be related to similar topics, while their diversity indicates the likely association with different topics. They developed the metric Laplacian score to rank topics, reflecting the degree of its discriminatory documents, in order to find the topics with high levels of discrimination, as well as the paired mutual information that calculates the information that two topics have in common. In this way, the similarity of topics can be measured, which can be used to maximize the diversity of topics and minimize redundancy. With the similar aim of measuring how different the topics are, \cite{9_wang_topic_over} applied the "Non-Markov Continuous-Time Model of Topical Trends" to calculate the average distance of word distributions between all topic pairs. As the authors themselves claimed, this method of calculating topic similarity is better suited to reduce topic redundancy. The study of \cite{10_mehta_evaluating_topic} proposed a method for evaluating the quality of each topic, using the metric of the silhouette coefficient. Using the latent topic model (LDA), the approach is based on clustering topics and using the silhouette index, which is often used to characterize the quality of elements in a cluster. Topics from multiple models are clustered based on the similarity of their word distributions. The quality of learned clusters is examined to distinguish weak topics from strong topics. In this work the clusters (e.g. weak and strong) are left unchallenged without further measuring the correlation of these clusters e.g. to investigate Euclidean distances in the metrics.

\section{Methodology}\label{sec:methodology}
\subsection{Data}\label{subsec:data}
We have considered different two-class text data sets. The first data set consists of short texts downloaded from the repository Stack Overflow\footnote{\url{https://archive.org/details/stackexchange}}. Applying the trust model proposed by \cite{11_alqundus_calculating_trust} \cite{12_alqundus_investigated_the_effect} the data is classified into four classes: \textit{very-trusted} (844 entries), \textit{very-untrusted} (117 entries), \textit{trusted} (904 entries) and \textit{untrusted} (347 entries). As we consider two-class data, we divided the data into the binary set \textit{trusted} / \textit{untrusted}, and \textit{very-trusted} / \textit{very-untrusted} documents. The second data set is created from two sources -Human-Aids and Mouse Cancer with 150 instances downloaded from PubMed. For simplicity, we will refer to this data as Aids vs Cancer.
A pre-processing procedure is applied on the data in order to transform it into a vector space data that could be subject to our algorithm.

\subsection{Pre-Processing}\label{subsec:preprocessing} 
A pre-processing step is necessary to convert the texts into vector space representations. We have used Knime \cite{14_berthold_the_konstanz_information} workflows for text preprocessing. Firstly, we perform cleaning of the text data using the following procedure: Punctuation Erasure, N-chars Filter, Number Filter, Case Converter (lower case), Stop-words Filter, Snowball Stemmer and Term Filtering. We used a language detector provided by Tika-collection to process English texts only. In the next step, the friction words are used as a dictionary that represents each document. These dictionaries are called bag-of-words (BoW) representation. BoW can be represented by Term-Frequency (TF) or binary. In the TF format counts the number of times a word appears in the document while the binary representation only distinguishes between 1 if the word is present and 0 otherwise. The number of words/features after performing the pre-processing stage is 714 for trusted vs untrusted data set, 848 words for the very-trusted vs very-untrusted data set and 1440 words for the human-aids vs mouse-cancer data set. For more detail see \cite{13_alqundus_exploring_the_impact}.

\subsection{Topic Clustering}\label{subsec:topicclustering}
The distance metric for clusters is a very important parameter for different clustering approaches. We use an agglomerative approach to hierarchical clustering. During agglomerative clustering, we merge the closest clusters as defined by the distance metric chosen. For computing the distance of two clusters, there is a variety of possible metrics, in this study we will use the four most popular: single-linkage, complete-linkage, average-linkage, and centroid-linkage. 

Single-link distance defines the distance between two clusters as the minimum distance between their members or is the distance between the nearest neighbors: $d_1(c_1,c_2)=\min\limits_{x\in c_1,y\in c_2}||x-y||$

It is called “single link” because it defines clusters that are close, if they have even a single pair of close points. Complete-link distance is defined as the distance between clusters. It is the maximum distance between the points of the clusters, or it is the distance between farthest neighbors: $d_2(c_1,c_2)=\max\limits_{x\in c_1,y\in c_2}||x-y||$

Average-linkage is the distance between each pair of points. In each cluster those are added up and divided by the number of pairs, which results in average inter-cluster distance. 
$d_3(c_1,c_2)=\frac{1}{n_{c_{1}} n_{c_{2}}} \sum_{i=1}^{n_{c_{1}}} \sum_{j=1}^{n_{c_{2}}} d(x_{i_{c{1}}},x_{j_{c{2}}})$
Centroid-linkage is the distance between the centroids of two clusters. $d_4(c_1,c_2)=||\bar{x_{c_{1}}}-\bar{x_{c_{2}}}||$
,where $\bar{x_{c_{1}}}$ is the centroid for the cluster $c_{1}$, while $\bar{x_{c_{2}}}$ is the centroid for cluster $c_{2}$.

\subsection{TopicsRanksDC Algorithm}
\begin{figure}
\vspace{-4mm}%to reduce space
\includegraphics[width=\textwidth]{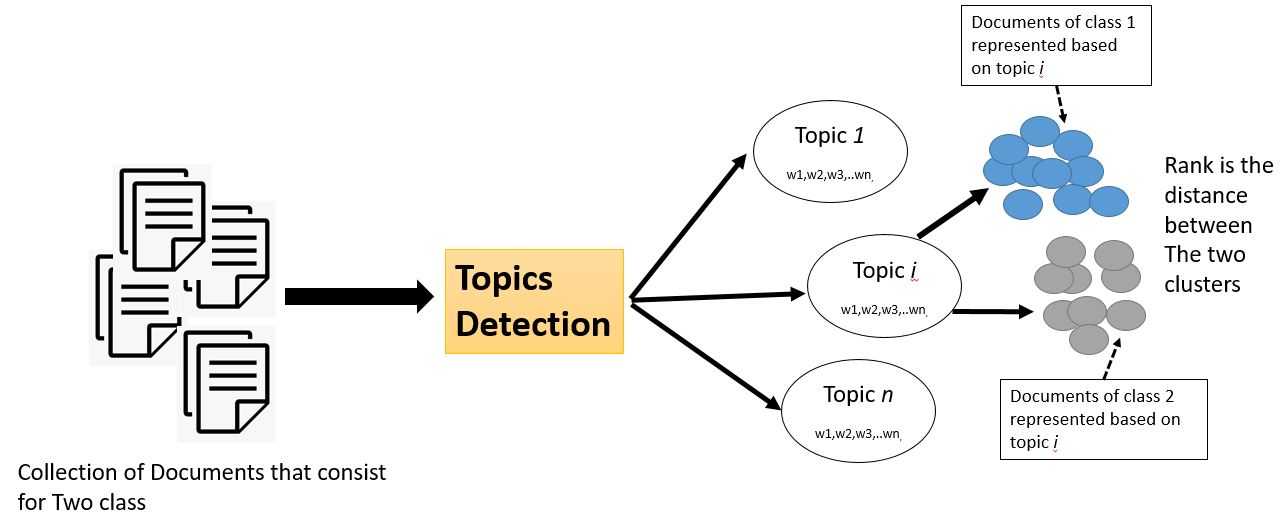}
\caption{The main workflow of TopicsRanksDC tool. The input is a collection of text documents belongs to two-class. Next is LDA or other approach for detecting topics. The last stage is ranking/scoring the topics on two-class data by distance of two clusters.} \label{fig:overview}
\vspace{-4mm}%to reduce space
\end{figure}

As illustrated in Figure \ref{fig:overview}, the input to the algorithm is a collection of text documents that we assume to be of two classes.  We mean by two-class data where part of the examples belongs to one label while the other belongs to the second label (documents about cancer patients vs documents about healthy).
The algorithm consists of two stages; the first stage detects the topics using LDA using the whole data (all the collection of text). A topic is a bag of words. The second stage is ranking or scoring each topic based on the distance of two clusters, where each cluster represents examples belongs to one label based only on the words of the given topic (Figure \ref{fig:overview}). The ranks stage actually defines the significance of each word in a topic in terms of separating the two-classes. In other words, suppose one has thousands of words that represent some data, and one wants to find out which groups of words (topics) were suitable to separate the two-classes. 
Let assume that we have detected n topics. Each topic contains k words. In order to calculate the rank or the significance of each topic related to the two classes we perform the following algorithm:
\begin{lstlisting}
For each topic i (i=1,..,n)  perform:
    a)	Create two clusters c1 and c2 of points that are
    represented by the words belonging to topic i. c1 
    contains the points of the first class (positive) and 
    c2 contains the points of the second class (negative).
    b)	Calculate four distance metrics, Single Linkage, 
    Complete Linkage, Average Linkage and distance between 
    the two centers so-called "centroids".
\end{lstlisting}
The rank we calculate gives an indication of how important the topic is for the separation of the two given classes, considering only the words associated with the specific topic. If the rank is close to zero, this implies that the two clusters are inseparable and the topic is not important to distinguish the two classes.
\begin{figure}
\vspace{-4mm}%to reduce space
\includegraphics[width=\textwidth]{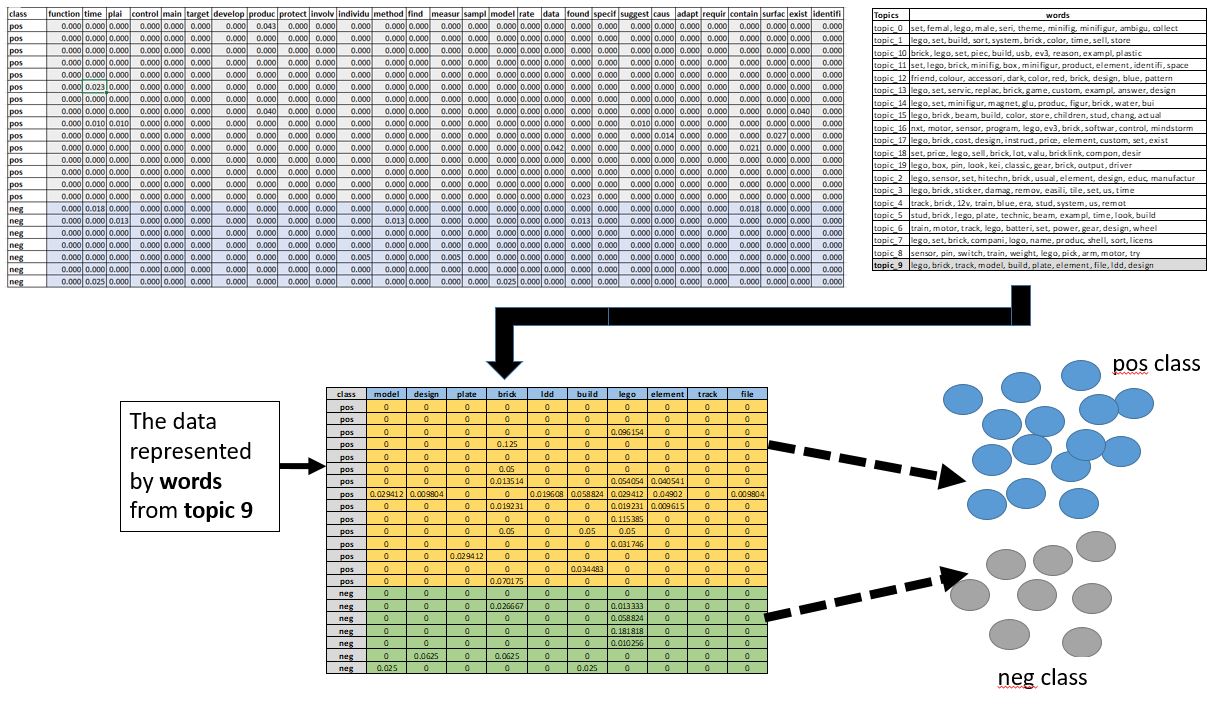}
\caption{illustrates of how the data is represented by words that belong to a specific topic creating new data with just columns that are associated with the words. Rows are the sample. The new data appears in the lower side of the figure is the two clusters.} \label{fig:exmapletopic}
\vspace{-4mm}%to reduce space
\end{figure}

The data in Figure \ref{fig:exmapletopic}, consists of columns that contain the features/words, while the rows are associated with documents. The columns class represents the labels for each document. In our case we assume two-class data sets (positive vs negatives). The new data appears on the lower part of the figure and is represented by the words belonging to topic number 9. The topics and their word lists appear on the upper part on the right side. The two clouds or clusters represent the new data in two- dimensional space. The aim is to calculate the distance between these two clusters. If the clusters are separable then the score/rank should be high, while if the two clusters are non- separable, the score should be close to zero indicating that the topic words are insignificant for the two-classes.

\section{Results and Discussion}\label{sec:results}
In order to have an idea about the two-classes, we have evaluated the performance of Random Forest (RF) on each data. The classifiers were trained and tested, with the division into 90\% training data and 10\% test data from the data generated by the pre-processing phase. The trusted/untrusted data sets used by the classifier are imbalanced, which can influence the classifier to the advantage of the set with more samples and is so called the problem of the imbalanced class distribution. We have applied an under-sampling approach that reduces the number of samples of the majority class to the minority class, thus reducing the bias in the size distribution of the data subsets. For more details see \cite{13_alqundus_exploring_the_impact}. We have set the ratio of the reduction to be up to 2 fold. We applied 100-fold Monte Carlo Cross Validation (MCCV)\cite{15_xu_monte_carlo}. The average of the 100 iteration is calculated to form as the final performance evaluation. Table \ref{tab1}, presents the results of RF. It is clear that the data of Aids | Cancer is much more separable than the data of trusted | untrusted. The result here is with considering all the features generated by the pre-process step to bring the raw data into vector space.
\begin{table}
\vspace{-4mm}%to reduce space
\centering
\caption{Classifier Performance of the Random Forest based on Bag-of- Words Model.}\label{tab1}
\begin{tabular}{|l|l|l|l|l|}
\hline
 data set &  Sen & Spe & F1 & Acc\\
\hline
vt vs vu	&0.95&	0.72&	0.84&	0.76\\
\hline
t vs u &	0.98&	0.85&	0.80&	0.69\\
\hline
Aids vs   Cancer&	0.98&	0.93&	0.96&	0.96\\
\hline
\end{tabular}
\begin{tablenotes}
      \small
      \item {vt=very trusted, t=trusted, u= untrusted, vu=very untrusted. Sen is sensitivity, Spe is specificity, F1 is F1 measure and Acc is accuracy}
\end{tablenotes}
\vspace{-4mm}%to reduce space
\end{table}
In order to test the algorithm TopicsRanksDC, we set the number of topics to be 10. Then we generate different size of topics. Size is the number of words in each topic. For each one of these options, we compute the significant/ranks of each topic. Table \ref{tab2} illustrates one sample of the output of the tool.

\begin{table}
\vspace{-4mm}%to reduce space
\centering
\caption{TopicsRnaksDC result on the data \textit{t} vs. \textit{u}}\label{tab2}
\begin{tabular}{|l|l|l|l|l|l|}
\hline
 cluster&	Topic Terms&	CentroidD&	MinD&	MaxD&	MeanD\\
\hline
topic\_5&	lego, piec, set, brick, model&	0.10&	0.00&	2.24&	1.18\\
\hline
topic\_6&	lego, set, brick, piec, box&	0.08&	0.00&	2.24&	1.16\\
\hline
topic\_1&	set, lego, http, list, bricklink&	0.09&	0.00&	2.24&	1.04\\
\hline
topic\_7&	brick, lego, stud, plate, piec&	0.08&	0.00&	2.24&	1.02\\
\hline
topic\_0&	brick, lego, motor, gear, batteri&	0.09&	0.00&	2.24&	0.91\\
\hline
topic\_9&	lego, block, water, duplo, set&	0.07&	0.00&	2.24&	0.90\\
\hline
topic\_8&	motor, lego, power, gear, us&	0.09&	0.00&	2.24&	0.88\\
\hline
topic\_4&	nxt, program, sensor, ev3, lego&	0.04&	0.00&	2.24&	0.81\\
\hline
topic\_3&	unit, pod, set, oppon, build&	0.06&	0.00&	1.73&	0.59\\
\hline
topic\_2&	friend, accessori, brick, hole, print&	0.06&	0.00&	2.00&	0.52\\
\hline
\end{tabular}
\begin{tablenotes}
      \small
      \item {Table \ref{tab2} shows the result of applying TopicsRnaksDC on the data trusted vs untrusted (t vs u). The number of LDA topics is 10 with 5 words in each topic. CentroidD is the Centroid-linkage, MinD is the Single-link distance, MaxD is the Complete-link distance, while MeanD is the average-link}
\end{tablenotes}
\vspace{-4mm}%to reduce space
\end{table}
To add sense of these values, we have generated random topics that include random words from the whole sets. The random topics used as input to the tool with the input of the algorithm binary data. The results of this experiment are illustrated in Table \ref{tab3}.
\begin{table}
\vspace{-4mm}%to reduce space
\centering
\caption{TopicsRnaksDC result on words in each topic selected randomly}\label{tab3}
\begin{tabular}{|l|l|l|l|l|l|}
\hline
 cluster&	Topic Terms&	CentroidD&	MinD&	MaxD&	MeanD\\
\hline
topic\_3&	happen, seri, fit, tell, avail&	0.02&	0.00&	2.00&	0.31\\
\hline
topic\_9&	get, prefer, fan, question, file&	0.03&	0.00&	2.00&	0.29\\
\hline
topic\_6&	tend, floor, probabl, mean, stand&	0.05&	0.00&	2.24&	0.29\\
\hline
topic\_7&	usual, easi, etc, structur, heavi&	0.05&	0.00&	2.00&	0.25\\
\hline
topic\_8&	half, take, stack, coupl, recommend&	0.04&	0.00&	1.73&	0.25\\
\hline
topic\_4&	stick, seller, steer, mindstorm, suspect&	0.06&	0.00&	2.00&	0.22\\
\hline
topic\_1&	war, car, guid, access, stop&	0.02&	0.00&	1.73&	0.17\\
\hline
topic\_2&	speed, calcul, shape, addit, save&	0.02&	0.00&	2.00&	0.16\\
\hline
topic\_5&	assembli, techniqu, regard, environ, displai&	0.01&	0.00&	2.00&	0.14\\
\hline
topic\_0&	place, true, imagin, vertic, particular&	0.02&	0.00&	1.73&	0.13\\
\hline
\end{tabular}
\begin{tablenotes}
      \small
      \item {Table \ref{tab3} represents the result of applying TopicsRnaksDC on the data trusted vs untrusted (t vs u.) The number of topics is 10 with 5 words in each topic selected randomly form all the words.}
\end{tablenotes}
\end{table}

\begin{figure}
\vspace{-2mm}%to reduce space
     \centering
\includegraphics[scale=0.8]{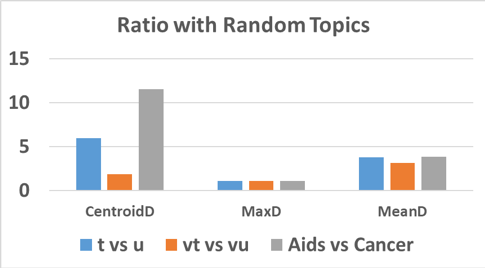}
\caption{Destruction of ratio with the random topics.} \label{fig3}
\end{figure}

Table \ref{tab4}  presents the ratio ranks between the topics generated by LDA and the random topics. The MeanD (average-link distance or d3 metric) shows ratio of above 3.6 for all the topics.
Figure \ref{fig3} summarizes the results ratio of TopicsRanksDC output applied on LDA topics and random topics. The topic 10 with words 5 is considered. Generally, Figure 3 shows that the model of TopicsRanksDC performs well and is much higher than the random topics. For example, for the Aids vs Cancer data the centroids of the two-classes represented by just the top ranked topics is very large. These results match the higher accuracy when using random forest.
\begin{table}
\vspace{-4mm}%to reduce space
\centering
\caption{Ratio ranks between the topics generated by LDA and the random topics.}\label{tab4}
\begin{tabular}{|l|l|l|}
\hline
 CentroidD&	MaxD&	MeanD\\
\hline
5.92&	1.12&	3.78\\\hline
2.80&	1.12&	4.01\\\hline
1.71&	1.00&	3.59\\\hline
1.54&	1.12&	3.99\\\hline
2.13&	1.29&	3.66\\\hline
1.08&	1.12&	4.17\\\hline
4.25&	1.29&	5.03\\\hline
1.61&	1.12&	4.96\\\hline
5.25&	0.87&	4.23\\\hline
2.76&	1.15&	4.05\\
\hline
\end{tabular}
\begin{tablenotes}
      \small
      \item {Table \ref{tab4} shows the ratio between values of LDA to random topics. The MinD column is discared is all is zero for all topics.}
\end{tablenotes}
\vspace{-4mm}%to reduce space
\end{table}

We tested TopicsRanksDC on fixed number of topics with variety of number of words. The 10 topics were generated by LDA with different count of words in the topic models. We consider 5, 10, 20 and 40 words as the size of topic models.
\begin{figure}
     \centering
%     \begin{subfigure}[b]{0.3\textwidth}
%         \centering
%         \includegraphics[width=\textwidth]{fig3.png}
%         \caption{Destruction of ratio with the random topics.}
%         \label{fig:y equals x}
%     \end{subfigure}
%     \hfill
     \begin{subfigure}[b]{0.48\textwidth}
         \centering
         \includegraphics[width=\textwidth]{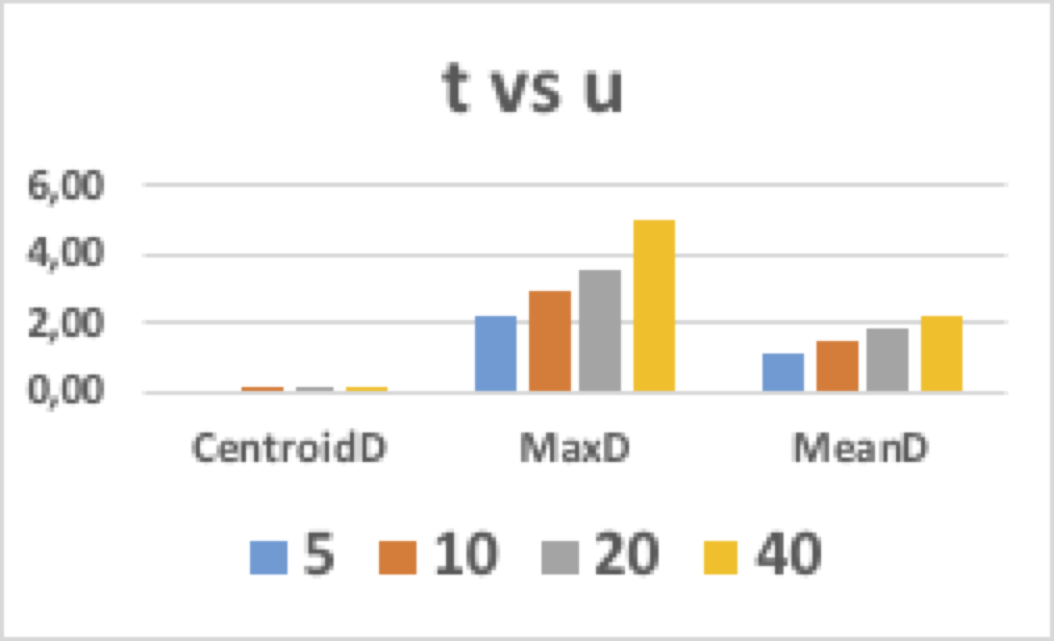}
         \caption{The distribution of the 3 distance metrics with different number of words in 10 topics. We remove the MinD as it zero for all words. The data is t vs u .}
         \label{fig4}
     \end{subfigure}
     \hfill
     \begin{subfigure}[b]{0.48\textwidth}
         \centering
         \includegraphics[width=\textwidth]{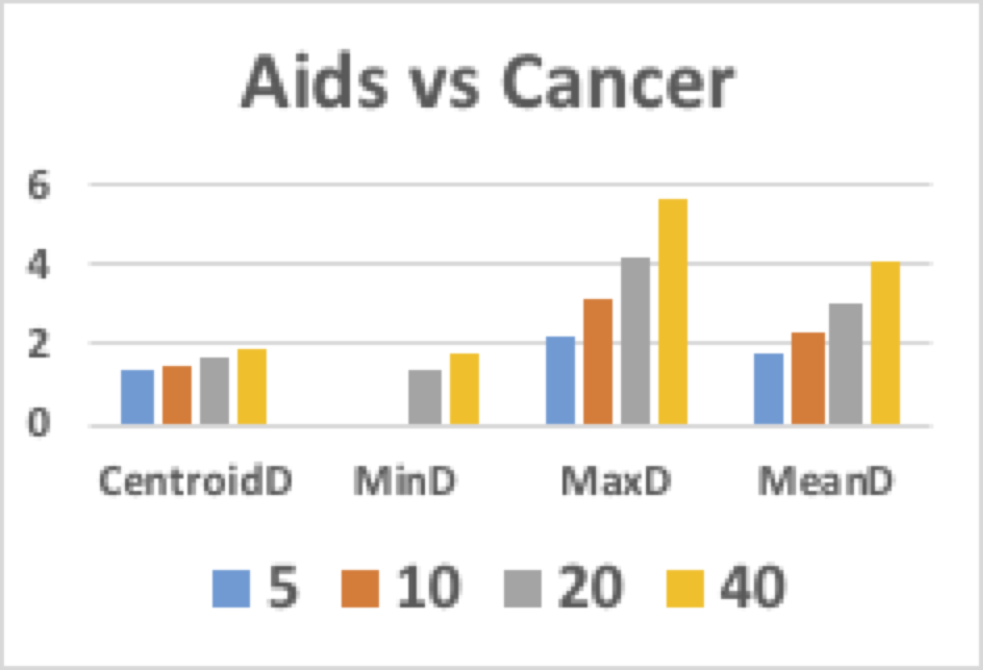}
         \caption{The distribution of the 4 distance metrics with different number of words in 10 topics. The data Aids vs Cancer.}
         \label{fig5}
     \end{subfigure}
        \caption{}
        \label{fig:three graphs}
\vspace{1mm}%to reduce space
\end{figure}

Figure \ref{fig4} and Figure \ref{fig5} show the influence of the size of the words (number of topics is fixed to 10). In Figure 4 that related to the t vs u data, we removed the results for MinD as all zeros, while it is clear that MaxD metrics are increasing significantly as the number of words is increasing and a slight increase is in the MeanD metric. The CentroidD metric is almost similar. Similar observations for the MaxD metric are presented in Figure \ref{fig5} for the Aids vs Cancer data. Additionally, the MinD is getting increase indicating that the data is more separable than the t vs u data. Also, the MeanD metric is increasing significantly.

\section{Conclusion and Future Work}
This work introduced a novel approach for topic ranking or scoring applied on two-class data sets. The score is actually the significance of the topic (set of words) for separating the two-classes. The scoring function used for computing the ranks is based on the distance between clusters associated with one of the classes represented by the words belong to a specific topic. Interestingly also simple metrics were successful in ranking the topics comparing to random topics. However, we assume that more metrics should be examined and as a future work, we will consider to use the machine learning approaches for ranking topics. One approach is considering the one-class classifier applied on text data \cite{16_manevitz_one_class_document} \cite{17_manevitz_one_class_svms}.

\end{document}